\documentclass[12pt,aps]{revtex4}
\usepackage{epsfig,amsmath,amssymb}

\newcommand{\fr}[1]{
             \frac{#1}}
\newcommand{\bea}{\begin{eqnarray}}
\newcommand{\eea}{\end{eqnarray}}

\newcommand{\ket}{{\rangle}}

\newcommand{\bra}{{\langle}}
\newcommand{\gc}{\bra\fr{\alpha_s}{\pi}G^2\ket}

\newcommand{\ga}{g_{{\cal A}}}

\newcommand{\dbar}{d\hskip -0.4em ^-}

\newcommand{\mc}[1]{{\mathcal{#1}}}
\begin{document}

\title{On  the color suppressed contribution to 
  $\overline{B_{d}^0} \rightarrow \, \pi^0 \pi^{0}$} 

\vspace{0.5cm}


\author{Jan O. Eeg and  Teresa Palmer}
\affiliation{Department of Physics, University of Oslo,
P.O.Box 1048 Blindern, N-0316 Oslo, Norway}

\vspace{0.5 cm}

\begin{abstract}

The  decay modes of the type 
$B \rightarrow \pi \, \pi $ are dynamically different. 
For the case  $\overline{B_{d}^0} \rightarrow \, \pi^+ \pi^- $
there is a substantial factorized contribution which dominates.
In contrast,  the decay mode 
  $\overline{B_{d}^0} \rightarrow \, \pi^0 \pi^{0} $ 
has a small  factorized contribution, being proportional to a  small Wilson
  coefficient combination. However,
for  the decay mode
  $\overline{B_{d}^0} \rightarrow \, \pi^0 \pi^{0} $
there is a  sizeable  nonfactorizable (color suppressed) contribution due to 
soft (long distance) interactions,
which
dominate the amplitude.

We estimate the branching ratio for the
 mode $\overline{B_{d}^0} \rightarrow \, \pi^0 \pi^{0} $ 
 in the heavy quark limit for the  $b$- 
quark.  In order to estimate color suppressed
 contributions we treat the energetic light ($u,d,s$) quark 
 within a variant of
 Large Energy Effective Theory   combined  with a recent extension
 of chiral quark models in terms of model- dependent gluon condensates.

We find that our  calculated color suppressed  amplitude is suppressed
 by a factor of order
$\Lambda_{QCD}/m_b$ with respect to the factorizable amplitude, as it should
according to QCD-factorization.
 Further, for 
 reasonable values of the constituent quark mass 
and the gluon condensate,  the calculated  nonfactorizable amplitude for 
  $\overline{B_{d}^0} \rightarrow \, \pi^0 \pi^{0} $ can easily accomodate 
the  experimental value. 
Unfortunately, the color suppressed  amplitude is very sensitive to  
the  values of these model dependent parameters. Therefore fine-tuning is
 necessary in order to obtain an amplitude compatible with the experimental
 result for $\overline{B_{d}^0} \rightarrow \, \pi^0 \pi^{0} $.
 A possible link to the triangle anomaly is discussed.

\end{abstract}

\maketitle

\vspace{1cm}

Keywords:
$B$-decays, factorization, gluon condensate. \\
PACS:  13.20.Hw ,  12.39.St , 12.39.Fe ,  12.39.Hg.

\newpage

\section{Introduction}

Due to numerous experimental results coming from BaBar and Belle, 
there is presently great interest in decays of $B$-mesons. 
 LHC will also provide us with more  data for such processes. 
$B$-decays of the type $B \rightarrow \pi \pi$ and 
$B \rightarrow K \pi$,  
where  the energy release is big  compared to the light
meson masses (heavy to light transitions),  has been treated within
{\it QCD factorization} \cite{BBNS} and
{\it Soft Collinear Effective  Theory} (SCET) \cite{SCET}. In 
 the high energy limit,  
 the amplitudes for such decay modes factorize into  products
of two matrix elements of weak currents, 
and  some nonfactorizable corrections of order $\alpha_s$
  can be calculated perturbatively.
However, there are additional contributions of order $\Lambda_{QCD}/m_b$
which cannot be reliably calculated within perturbative theory \cite{BBNS}.
The  so called pQCD-model and QCD sum rules have  also been used
 for $B$-meson decays
\cite{pQCD,KMMM}.

For decay modes which are of the heavy to heavy type,
involving $b$- and   $c$-quarks, the decay amplitudes
have  been described within {\it Heavy Quark Effective Field Theory}
(HQEFT) \cite{neu}.
Some transitions of {\em heavy to heavy} type in the heavy quark 
limits $(1/m_b) \rightarrow 0$ like $B - \bar{B}$ mixing \cite{ahjoeB}
has been studied within
{\it Heavy Light Chiral Perturbation Theory} (HL$\chi$PT)
\cite{itchpt}.
Furthermore, other  transitions which are formally {\em heavy to heavy} 
 in the heavy quark 
limits $(1/m_b) \rightarrow 0 $ {\it and  } $(1/m_c) \rightarrow 0$, 
like the Isgur-Wise function \cite{IW} for  $B \rightarrow D$, 
have been studied within
HL$\chi$PT \cite{itchpt}.  The cases 
  $\bar{B} \rightarrow D \overline{D}$ \cite{EFHP} and
 $B \rightarrow D^* \gamma$ \cite{GriLe,MacDJoe}
 have also been studied within such a framework, even if the 
energy release in these processes is   above the chiral symmetry 
breaking scale.
 Still this framework  give  amplitudes of  the right order
 of magnitude.   
The calculation of such transitions have in addition been supplemented 
with calculations within a {\it Heavy Light Chiral Quark Model} (HL$\chi$QM) to
 determine quantities which are not determined within HL$\chi$PT itself
\cite{EHP,EFHP,MacDJoe}.

As pointed out in a series of papers \cite{EHP,EFHP,MacDJoe,LLJOE},
 there are processes
which have factorized amplitudes multiplied by a very small Wilson coefficient 
combination, such that nonfactorized amplitudes are expected to 
dominate. Examples are
$\overline{B_{d,s}^0} \rightarrow D^0 \, \overline{D^0}$ \cite{EFHP} , 
 $\overline{B^0} \rightarrow D^0 \, \eta'$ \cite{EHP} and 
 $\overline{B_d^0} \rightarrow D^0 \, \pi^0$.
 The latter process 
$\overline{B_d^0} \rightarrow D^0 \, \pi^0$ was considered 
recently \cite{LLJOE,LELeg}.
 In that case
 a heavy $b$-quark decaying to a light, but  energetic quark was involved.
Then the light energetic quark might be  described by an effective theory.
The first version of such a framework was {\it Large Energy Effective Theory}
 (LEET) \cite{GD,charles}.
The HQEFT covers processes where the heavy quarks carry the main
part of the momentum in each hadron.
 To describe processes where energetic light quarks emerge from decays
 of heavy $b$-quarks,  LEET
 was introduced \cite{GD} and used to study  
the current for $B \rightarrow \pi$ \cite{charles}.  

The idea was that LEET should do for energetic light quarks what 
 HQEFT did for heavy quarks.
In HQEFT one splits off the heavy motion
 from the full heavy quark field, 
thus obtaining a reduced field depending on the velocity of the heavy quark.
 Similarly, in LEET  one splits off the large energy from the full field of the
 energetic light quark,  thus obtaining an effective description
 for a reduced light 
quark which depends on a light-like four vector. 
 It was later shown that LEET in its
 initial formulation was incomplete and
 did not fully reproduce  QCD physics
\cite{Uglea}. Then  LEET was further developed
 to be fully consistent with QCD and 
became the Soft Collinear Effective Theory (SCET)  \cite{SCET}.

In the present  paper we consider  decay modes of the type 
$B \rightarrow  \pi \, \pi$.
The decay mode  $\overline{B_{d}^0} \rightarrow \, \pi^- \, \pi^{+}$
has a substantial factorized amplitude, given by the current matrix element
for $\overline{B_d^0} \rightarrow \pi^+$ transition times
the matrix element of the weak current for the outgoing $\pi^-$, which 
is proportional to  the pion decay
 constant $f_\pi$.
The relevant Wilson coefficient is also the maximum possible, namely of 
order one times
the  relevant Cabibbo-Kobayashi-Maskawa (CKM) quark mixing 
 factors and the Fermi coupling constant.
This is in contrast to the process 
$\overline{B_{d}^0} \rightarrow \, \pi^0 \pi^{0}$ which is color suppressed.
As said above, decays of the type $B \rightarrow 2 \pi$
  have been extensively  studied within
 QCD-factorization, SCET, and QCD sum rule methods \cite{Btotwopi}.
In spite of tremendous efforts it has not been possible to obtain an amplitude 
 compatible  with  
the experimental result
 $\overline{B_{d}^0} \rightarrow \, \pi^0 \pi^{0}$.
The purpose of this paper is  study this decay mode within an alternative model
 dependent  framework.

First we  point out that the factorized contribution to the decay mode
 $\overline{B_{d}^0} \rightarrow \, \pi^0 \pi^{0}$,
 which is  given by
the $B \rightarrow \pi$ transition amplitude times the decay constant of the 
$\pi^0$ meson, is almost zero because it is proportional to  a very small Wilson
coefficient combination. For the dominant nonfactorizable (color suppressed)
  amplitude for 
 $\overline{B_{d}^0} \rightarrow \, \pi^0 \pi^{0}$
we will, as mentioned above, use a  model
named {\it Large Energy Chiral Quark Model} (LE$\chi$QM) 
 recently constructed and used   
to handle the process
 $\overline{B_{d}^0} \rightarrow \, \pi^0 D^{0}$  \cite{LELeg,LLJOE}.
Here  a variant of LEET was combined with ideas from previous chiral
 quark model ($\chi$QM) calculations similarly to what 
 has been done for other nonleptonic decays
\cite{ahjoe,ahjoeB,EHP,MacDJoe,BEF}.

A priori it might look strange to use the framework of chiral quark models 
when the energy release is big compared to the  chiral symmetry breaking
 scale $\Lambda_\chi$. The point is that 
the motion of the heavy quark or energetic light quark can be split off, and 
the various versions of heavy-light or large energy chiral quark models 
 and a corresponding chiral perturbation theory ($\chi PT$)
 can be used to describe the 
redundant strong interactions 
corresponding to momenta of order 1 GeV and below.

It might be argued that we should have used the full SCET theory as 
the basis  our new model. However, the purpose of our paper is to estimate,
 in analogy with   previous papers
 \cite{ahjoeB,EHFiz,EHP,MacDJoe,ahjoe,EHbeta,BEF,epb},
 the effects of
 soft gluon emission in terms of gluon condensates, where transverse 
quark momenta and   collinear
 gluons will not play an essential  role. In any case this
 construction \cite{LLJOE} will be a model. 
Therefore it suffices for our purpose to use the more simple formulation of
 LEET.  We will combine LEET with chiral quark models ($\chi$QM)
\cite{chiqm,pider,BHitE,BijRZAn,EHFiz}, containing only soft gluons 
making condensates. 
 In LE$\chi$QM \cite{LLJOE} an energetic quark is bound to a soft quark
with an a priori unknown coupling,  as proposed in \cite{EHFiz}. The
unknown coupling is determined by calculating the  known 
$B \rightarrow \pi$ current matrix element within the model \cite{LLJOE}.
 This  fixes 
  the unknown  coupling because the matrix element of this current is known
\cite{charles}.
   Then, in the next step, we use this coupling  
 to  calculate the nonfactorized (color suppressed) 
amplitude contribution to  
 $\overline{B_{d}^0} \rightarrow \, \pi^0 \pi^{0}$ in terms of the lowest
 dimension gluon condensate, as have been done for other nonleptonic decays
\cite{ahjoe,ahjoeB,EHP,MacDJoe,BEF}. 
After the quarks have been
integrated out, we obtain an effective theory containing 
 soft light mesons as in HL$\chi$PT, but  also
 fields describing energetic light mesons. A similar idea with a 
combination of SCET  with 
HL$\chi$PT  is  considered in \cite{GrinPir}.
The LE$\chi$QM was constructed in analogy with the previous
 {\it Heavy Light Chiral Quark Model} (HL$\chi$QM)\cite{ahjoe}
 and may be considered to be an extension of that model.

One might think that to be completely consistent,  we should also have
 calculated the Wilson coefficients 
 within a relevant large energy framework. 
For this purpose the use of LEET would be dubious because it is an 
incomplete theory
as mentioned above. However, as we will see below, the main uncertainty in
our final amplitude will be due to uncertainty in our model dependent 
 gluon condensate due to emission of soft gluons.
Therefore  the  Wilson coefficients calculated  within full
 QCD as in \cite{QCDloop} will be appropriate for our purpose.

In the next section (II) we present the
 weak four quark Lagrangian and its factorized and nonfactorizable
 matrix elements.
In section III we present our version of LEET,
and in section IV we present the new model LE$\chi$QM
to include energetic light quarks and mesons.
In section V we  calculate the 
nonfactorizable matrix elements due to soft gluons expressed through the (model
dependent) quark condensate. 
 In section VI we give the results and 
conclusion.

\section{The effective Lagrangian at quark level}

We will study  decays  of $\overline{B^0_d}$ 
 generated  by the weak quark process
 $b \rightarrow u \bar{u}d$. We restrict ourselves to
processes  where the $b$-quark decays. This means  
 the quark level  processes  
$b  \rightarrow   d u \bar{u} \, $.
Processes where the anti- $b$-quark decays proceed analogously.
The  effective weak Lagrangian at quark level is \cite{QCDloop}
(neglecting penguin operators)
\begin{equation}
  {\cal L}_\mathrm{eff} = -
  \frac{G_F}{\sqrt{2}}V_\textrm{ub}V^*_\textrm{ud}\left[ c_A \, Q_A
 + c_B \,  Q_B \right],
\label{effLag}
\end{equation}
where the subscript $L$ denotes the left-handed fields: $q_L \equiv L \, q$,
 where $L \equiv (1\, - \, \gamma_5)/2$ is the left-handed projector 
in Dirac-space. 
The local operator products $Q_{A,B}$ are defined as
\begin{equation}
  Q_A = 4 \, \bar{u}_L \gamma_\mu  b_L \; \bar{d}_L \gamma^\mu  u_L
\quad ; \; \, 
  Q_B = 4 \, \bar{u}_L \gamma_\mu  u_L \;  \bar{d}_L \gamma^\mu  b_L .
\end{equation}
In these operators summation over color is implied. In Eq.~(\ref{effLag}),
 $c_A$ and $c_B$ are Wilson coefficients. At tree level $c_A=1$ and
$c_B=0$. At one loop level, a contribution to $c_B$ is also generated, and 
$c_A$ is slightly increased. These effects are handled in terms of 
the {\it Renormalization Group Equations} (RGE)\cite{QCDloop}, and the
 coefficients can be calculated at for instance $\mu =m_b$ or $\mu$= 1 GeV.
Using the color matrix identity 
\begin{equation}
 2 \; t_{i n}^a \; t_{l j}^a \, = \, 
\delta_{i j}\delta_{l n}  \, - 
\,    \frac{1}{N_c} \delta_{i n} \delta_{l j} \; \; ,
\label{eq:color-identity}
\nonumber
\end{equation} 
 and Fierz rearrangement, 
the amplitudes for the  processes
 $\overline{B^0} \rightarrow \pi^+ \pi^-$ may be written as
\begin{eqnarray}
   \mathcal{M}_{\pi^+\pi^-} \, =
\, 4 \, \frac{G_F}{\sqrt{2}}V_\textrm{ub}V^*_\textrm{ud}
 & \left[
     \left(c_A+\frac{1}{N_\textrm{c}}c_B \right)
     \bra{\pi^-} | \bar{d}_L \gamma_\mu \, u_L | 0 \ket \bra{\pi^+} | \bar{u}_L
\gamma^\mu \, b_L |\bar{B}^0 \ket \right.   \nonumber  \\
       &\left.+  2 \, c_B \, \bra{\pi^+\pi^-} | \bar{d}_L \gamma_\mu t^a u_L 
\bar{u}_L \gamma^\mu t^a b_L  | B^0 \ket \right],
\label{BDpiFact}
\end{eqnarray}
and for   $\overline{B^0} \rightarrow \pi^0 \pi^0$
\begin{eqnarray}
    \mathcal{M}_{\pi^0\pi^0} = 4 \, \frac{G_F}{\sqrt{2}} 
V_\textrm{ub}V^*_\textrm{ud} & 
\left[ \left(c_B+\frac{1}{N_\textrm{c}}c_A \right)
     \bra{\pi^0} | \bar{u}_L \gamma_\mu u_L | 0 \ket \bra{\pi^0} | \bar{d}_L
\gamma^\mu b_L  | \bar{B}^0 \ket \right. \nonumber \\
       &\left.+  2 \, c_A \bra{\pi^0\pi^0} | \bar{d}_L 
\gamma_\mu t^a b_L \bar{u}_L \gamma^\mu t^a u_L | B^0 \ket
         \right] \, .
\label{BDpiNonFact}
\end{eqnarray}
Here the  terms proportional to $2 c_A$ and $2 c_B$ with color
 matrices inside the matrix elements are
 the genuinely nonfactorizable contributions.

Since $c_A$ is of order one and  $c_B$ of order $- 1/3 \,$ \cite{EHP,LLJOE} ,
 we refer to the coefficients 
 \begin{equation}
  c_{f} \, \equiv \, \left(c_A+\frac{1}{N_\textrm{c}}c_B\right) \, \simeq \,
  1.1   \qquad ; \quad
  c_{nf} \; \equiv \;  \left(c_B+\frac{1}{N_\textrm{c}}c_A \right) \, \simeq
  \, 0 \; \, ,
\end{equation}
as favorable ($c_f$) and nonfavorable ($c_{nf}$) coefficients, respectively.
Thus, the decay mode
 $\overline{B^0_d} 
\rightarrow  \pi^+\pi^-$ has a sizeable factorized amplitude
proportional to $c_f$. In contrast, the decay mode
$\overline{B^0_d} \rightarrow \pi^0\pi^0$ has a factorized amplitude
proportional to the  nonfavorable coefficient $c_{nf}$ which is close to
zero. In this case we expect the nonfactorizable term (involving
colour matrices) proportional to $2 c_A$ to be dominant, i.e.  
the last line of eq. (\ref{BDpiNonFact}) dominates.
A substantial part 
 of this paper is dedicated to the calculation of this nonfactorizable
contribution to the $\overline{B^0_d} \rightarrow \pi^0 \pi^0$ decay amplitude.

Thus the main task of this paper will be to calculate the matrix 
element of the operator $Q_C$ consisting of the product of two colored currents
occurring in the last line of  eq. (\ref{BDpiNonFact}):
\begin{eqnarray}
Q_C \; = \;  \left(\bar{d}_L \gamma_\mu t^a b_L \right) \,
 \left( \bar{u}_L \gamma^\mu t^a u_L \right) 
\label{Colop}
\end{eqnarray}
for the color suppressed process $\overline{B^0_d} \rightarrow \pi^0 \pi^0$. 
This matrix element  will be estimated in section~V where we use 
the LE$\chi$QM to estimate nonfactorizable amplitudes in terms of emission 
of soft gluons making gluon condensates.

\section{An  energetic light quark  effective description (LEET$\delta$)}
\label{LEET}

An energetic light quark might, similarly to a heavy quark,  carry
 practically  all the energy $E$ of
the meson it is a part of.
 The difference is that now   the mass of the energetic quark  is close to
 zero compared to the heavy quark mass $m_Q$ and $E$, which
 are assumed to be of the same 
order of magnitude. We  assume that the energetic light 
 quark is emerging from the decay of a
 heavy quark $Q$ with momentum $p_Q =  m_Q \, v \, + k$. The heavy quark
 is described by the HQEFT Lagrangian for  the reduced quark field $Q_v$ 
\cite{neu}: 
\begin{equation}
{\cal L}_\mathrm{HQET} = \bar{Q}_v \, (iv\cdot D) \, Q_v 
   + \mathcal{O}(1/m_Q) \; ,
\label{eq:LeffFINAL}
\end{equation}
where $Q_v$ is the reduced heavy quark field (often named $h_v$ in the 
literature), $v$ its four velocity and $m_Q$
the mass of the heavy quark.

The momentum of
 the light energetic light quark $q$ can be written
\begin{equation}
p^\mu_q = E \, n^\mu + k^\mu \quad ,   
\qquad \;   |k^\mu| \ll |E \, n^\mu| \quad , \qquad  m_q \ll E \; ,
\label{eq:pq}
\end{equation}
where $E$, which is of order $m_Q$, is the energy of the energetic light quark, 
$m_q$ is the light quark mass.
Further, 
 $n$ is the light-like four
 vector wich might be chosen to have 
the space part  along the z-axis,
 $n^\mu = (1; 0, 0, 1)$, 
 in the frame of the heavy quark where $v = (1,\underline{0})$.
Then  $(v\cdot n) = 1$ and  $n^2 \, = \, 0$.
Inserting this in the  regular quark propagator, 
In the limit where the approximations in (\ref{eq:pq}) are valid, 
 we obtain the propagator
\begin{equation}
 S(p_q) \, = \, 
\frac{\gamma \cdot p_q + m_q}{p_q^2 -m_q^2}  \; \;  \rightarrow \; \;
  \frac{\gamma \cdot n}{2 n \cdot k} \; \, .
\label{eq:LEETprop}
\end{equation}
This propagator is the starting point for the Large Effective Theory
(LEET) constructed
in Ref. \cite{charles}.

Unfortunately, the combination of LEET  with $\chi$QM will lead to
 infrared divergent loop integrals  for $n^2=0$ 
(see section \ref{HLET+LEET}). 
Therefore,   the formalism  was modified \cite{LELeg,LLJOE} and instead 
 of  $n^2=0$,
 we use $n^2 = \delta^2$, with $\delta = \nu/E$
where $\nu \sim \Lambda_{QCD}$,  such that $\delta \ll 1$. An expansion in 
$\delta$ will then within our model be equivalent to an expansion in 
$\Lambda_{QCD}/m_b$.

In the following we describe  the 
modified LEET \cite{charles} where we  keep $\delta \neq 0$
with $\delta \ll 1$. We call this construction LEET$\delta$ 
\cite{LLJOE} and 
 define  the  {\it almost} light -like vectors
\begin{equation}
 n = (1,0,0,+\eta),\qquad ; \quad 
 \tilde{n} = (1,0,0,-\eta),
\end{equation}
where
$\eta = \sqrt{1-\delta^2} $. This means that
\begin{eqnarray}
n^\mu + \tilde{n}^\mu = 2v^\mu \; \,  ,\, n^2 = \tilde{n}^2 = 
\delta^2 \, , \, 
v\cdot n = v\cdot \tilde{n} = 1 \; \; , \, n \cdot  \tilde{n} \, = 2 -
\delta^2 \, .
\end{eqnarray}
In the following we use the  projection
 operators given by
\begin{equation}
  \mc{P}_+ = \frac{1}{N^2}\gamma \cdot n(\gamma \cdot
  \tilde{n}+\delta) \quad , \; 
  \mc{P}_- = \frac{1}{N^2}( \gamma \cdot \tilde{n}-\delta)
\gamma \cdot  n  \; \, ,
 \label{eq:LEETprojectors}
\end{equation}
where  $N \, = \sqrt{2 \, n \cdot  \tilde{n}} \;
 =  \; 2 \, + {\cal O}(\delta^2)$ .
 One  factors out the main energy dependence, just as was analogously 
done  in HQEFT, and  define the projected reduced quark fields\cite{charles}   
\begin{equation}
  q_\pm(x) = e^{iEn\cdot x}\mc{P}_\pm q(x) \quad  , \; \;  
  q(x) = e^{-iEn\cdot x}\left[q_+(x) + q_-(x)\right].
 \label{eq:LEETqpm}
\end{equation}

As in  \cite{charles}, the field $q_-$ 
was eliminated and one obtained  for
 $q_+   \equiv q_n$ the effective Lagrangian \cite{LLJOE}:
\begin{eqnarray}
  {\cal L}_{LEET\delta} \, = \, 
 \bar{q}_n \left(\frac{\gamma \cdot  \tilde{n} + \delta}{N} \right)
(i n  \cdot D) q_n
  + \frac{1}{E}\bar{q}_n \, X \, q_n + \mathcal{O}(E^{-2}) \; ,
\label{eq:LEETdelta-Lag}
\end{eqnarray}
which (for $\delta = 0$) is  the first part of the SCET  Lagrangian.
The operator $X$  is given in \cite{LLJOE}.
Equation (\ref{eq:LEETdelta-Lag}) yields the LEET$\delta$
 quark propagator
\begin{equation}
S_n(k) \, = \, \mc{P}_+ \,  \left[\frac{\gamma \cdot \tilde{n} +
                   \delta}{N}(n \cdot k)\right]^{-1}
                   = \frac{\gamma \cdot n}{N(n\cdot k)} \; \; ,
\label{eq:LEETdelta-prop}
\end{equation}
which reduces to (\ref{eq:LEETprop}) 
 in the  limit $\delta\rightarrow 0$.
In addition, for a light energetic  quark, the 
 propagator within SCET  \cite{SCET} 
will for small transverse
 quark momenta  $p_\perp \, \rightarrow \, 0$  coincide with
Eq.  (\ref{eq:LEETdelta-prop}).

Based on LEET, it was found  \cite{charles}
in the formal limits $M_H \rightarrow \infty$ and 
$E\rightarrow\infty$, that a heavy $H \, = \, (B,D)$ meson decaying
by the weak hadronic vector current 
$V^\mu $  to a light pseudoscalar meson is described by a matrix element
$\langle P \, | \, V^\mu \, | \, H \rangle$   of the form
\begin{equation}
\langle P| V^\mu |H \rangle = 2E\left[\zeta^{(v)}(M_H,E) \, n^\mu 
                      + \zeta^{(v)}_1(M_H,E) \, v^\mu  \right] \; , 
\label{HECurrent}
\end{equation}
where
\begin{equation}
\zeta^{(v)}   = C \frac{\sqrt{M_H}}{E^2} \quad , 
\; \, C \sim (\Lambda_\textrm{QCD})^{3/2} \qquad , \quad 
\frac{\zeta^{(v)}_1}{\zeta^{(v)}} \sim \frac{1}{E} \; \; .
\label{eq:charlesEq}
\end{equation}
This behaviour is consistent  with the energetic quark
 having  
$x$  close to one, where $x$ is the quark 
 momentum fraction  of the outgoing pion \cite{charles}.

\section{Extended chiral quark model for heavy and 
 energetic light
  quarks (LE$\chi$QM)}
\label{HLET+LEET}

The  chiral quark model ($\chi$QM) \cite{chiqm,pider}  and the
 Heavy-Light Chiral Quark Model (HL$\chi$QM)
 \cite{ahjoe}, include meson-quark couplings and thereby
 allow us to calculate amplitudes and chiral  Lagrangians
 for processes involving heavy
 quarks and  low energy light
quarks. In this section we will extend these models to include also
 hard, energetic light quarks.

For the pure light and soft sector the $\chi$QM Lagrangian  can be written 
as \cite{chiqm,BEF}:
\begin{equation}
 {\cal L}_{\chi QM} = \bar{\chi} \left[ \gamma \cdot (i D + \mathcal{V})
       +\gamma \cdot \mathcal{A} - m \right]  \chi  \; ,
\label{constLagr}
\end{equation}
where $m$ is the  constituent mass term
being due to chiral symmetry breaking. The small current mass 
term is neglected here.
 Here we have 
 introduced the flavor rotated fields $\chi_{L,R}$: 
\begin{eqnarray}
                 \chi_L =  \xi^\dagger \, q_L \qquad , \quad 
                 \chi_R =  \xi  q_R \; ,
\label{xirot}
\end{eqnarray}
where $q$ is the light quark flavor triplet and:
\begin{eqnarray}
            \xi \,= \; \exp\{i \Pi/f\}  \qquad , \quad 
\Pi &=& \left(\begin{array}{ccc}
     \frac{\pi^{0}}{\sqrt{2}} + \frac{\eta}{\sqrt{6}} 
     & \pi^{+} & K^{+} \\
     \pi^{-} & -\frac{\pi^{0}}{\sqrt{2}} +
     \frac{\eta}{\sqrt{6}} & K^{0} \\
     K^{-} & \bar{K}^{0} & 
     -\frac{2\eta}{\sqrt{6}}\end{array}\right) \; .     
\label{xidef}
\end{eqnarray}
Further, 
$\mathcal{V}_\mu$ and $\mathcal{A}_\mu$ are vector and axial vector fields,
 given by
\begin{equation}
 \mathcal{V}_\mu \equiv \frac{i}{2}(\xi^\dagger \partial_\mu\xi +
 \xi\partial_\mu\xi^\dag) \quad , \quad 
 \mathcal{A}_\mu \equiv -\frac{i}{2}(\xi^\dagger \partial_\mu\xi -
 \xi\partial_\mu\xi^\dag) \; .
\end{equation}

To couple the heavy quarks to mesons there are additional meson-quark 
couplings within HL$\chi$QM \cite{ahjoe}:
\begin{equation}
{\cal L}_\mathrm{int} = -G_H \left[\bar{\chi}_a \, \bar{H}^a_v \, Q_v 
+ \bar{Q}_v \, H^a_v \, \chi_a\right]  \; ,
\label{HLInteract}
\end{equation}
where $Q_v$ is the (reduced) heavy quark field and  $H$ is the 
heavy $(0^-,1^-)$ meson field(s)
\begin{equation}
H^{(+)}_v = P_+(v) \left(\gamma\cdot P^* - i\gamma_5 \, 
P_5\right) \; ,
\end{equation}
$P^*_\mu$ being the $1^-$ and $P_5$ the $0^-$ fields,
and   $P_+(v) = (1 + \gamma \cdot v)/2$. The 
  quark-meson coupling $G_H$  
is  determined within the HL$\chi$QM to be \cite{ahjoe}
\begin{equation}
G_H^2 \; = \; \frac{2 m}{f_\pi^2} \, \rho \qquad , \quad 
\rho = \frac{(1+3g_A)}{4(1 + \frac{m^2N_c}{8 \pi f_\pi^2} \, - \,
 \frac{\eta_H}{2 m^2 \, f_\pi^2}  \langle\frac{\alpha_s}{\pi}G^2\rangle)} \; ,
\label{GHcoupling}
\end{equation}
where $\eta_H  = (8-\pi)/64$. The quantity $\rho$ is  of order one.

For hard light quarks and chiral quarks coupling
to  a hard light meson multiplet field $M$, one extends the ideas of 
$\chi$QM and HL$\chi$QM, and assume that the energetic light mesons 
couple to light quarks with a derivative coupling to an axial
 current \cite{LLJOE}: 
\begin{eqnarray}
{\cal L}_{\mathrm{int}q}
 \; \sim \;  
\bar{q} \, \gamma_\mu \gamma_5(i \, \partial^\mu M) \, q \; \,  .
\label{Ansatz}
\end{eqnarray}
One  combines  LEET$\delta$ with the $\chi$QM
and assume that the ingoing light quark and the outgoing meson are energetic
 and  have the behaviour $\exp{(\pm i E n \cdot x)}$ as in (\ref{eq:LEETqpm}).
To describe  (outgoing) light  energetic mesons, we use an octet 
$3\times 3$ matrix field $M =\exp{(+ i E n \cdot x)} \,  M_n$ of
 the same form as $\Pi$ in (\ref{xidef}):
\begin{eqnarray}
M_n &=& \left(\begin{array}{ccc}
     \frac{\pi^{0}_n}{\sqrt{2}} + \frac{\eta_n}{\sqrt{6}} 
     & \pi^{+}_n & K^{+}_n \\
     \pi^{-}_n & -\frac{\pi^{0}_n}{\sqrt{2}} +
     \frac{\eta_n}{\sqrt{6}} & K^{0}_n \\
     K^{-}_n & \bar{K}^{0}_n & 
     -\frac{2\eta_n}{\sqrt{6}}\end{array}\right) \; ,
\end{eqnarray}
where $\pi^{0}_n$, $\pi^{+}_n$, $K^{+}_n$ etc. 
are the  energetic light meson fields with momentum $\sim E n^\mu$. 

Combining  (\ref{Ansatz}) with the use of the rotated
 soft quark fields in (\ref{xirot}) and  using 
$\partial ^\mu \rightarrow i E \,  n^\mu$
 one  arrives at the ansatz for the LE$\chi$QM interaction Lagrangian:
\begin{eqnarray}
{\cal L}_{\mathrm{int}q \delta}
  \; = \; 
         \, G_A \, E 
\bar{\chi} \,  (\gamma \cdot n) \, Z \, q_n \, + \, h.c.
\; \, , 
\label{eq:LEETdHLciQM}
\end{eqnarray}
where $q_n$ represents an energetic light quark having momentum
fraction close to one and $\chi$ represents a soft quark
 (see Eq. (\ref{xirot})). Further, the coupling
$G_A$ is  determined  by physical 
requirements \cite{LLJOE,charles},
and 
\begin{eqnarray}
Z         = \xi M_R \, R - \xi^\dag M_L \, L \; \, .
\end{eqnarray}
Here $M_L$ and $M_R$ are both equal to $M_n$, but they have formally
different transformation properties, 
This is analogous to the use of
quark mass matrices  ${\cal M}_q$ and ${\cal M}_q^\dagger$  in standard
 {\it Chiral Perturbation Theory} ($\chi$PT). They are in practice equal,
 but have formally different transformation properties.

The axial vector coupling introduces a  factor $\gamma \cdot n$ to the vertex
(see (\ref{eq:LEETdHLciQM})), which  simplifies the Dirac algebra
 within the loop integrals.
\begin{figure}[t]
\begin{center}
   \epsfig{file=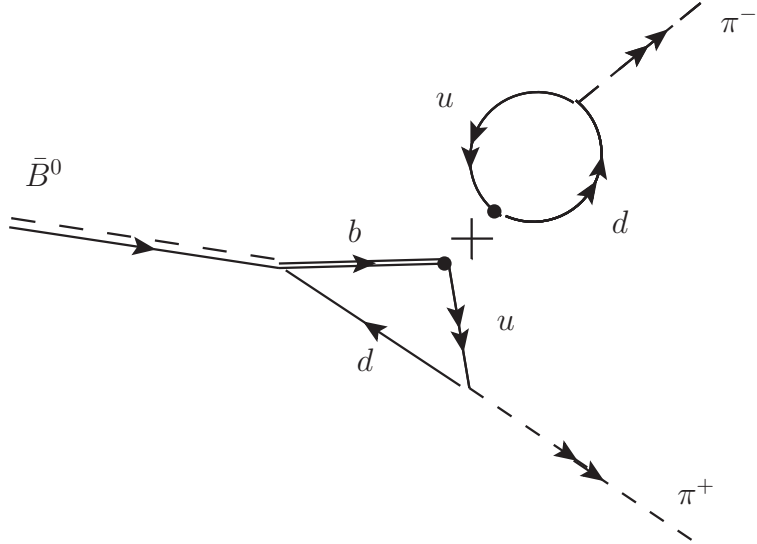,width=11cm}
\caption{The factorized contribution to the $B^0 \rightarrow \pi^+ \pi^-$ decay,
         as described in combined  HL$\chi$QM and LE$\chi$QM. Double lines, 
single lines and the single line with two arrows are representing heavy
 quarks, light soft quarks and light energetic quarks, respectively. 
Heavy mesons are represented by a sigle line combined with a parallel
 dashed line,
 and a light energetic pion is represented by a dashed line with double
 arrow.}
\label{fig:BDpiF}
\end{center}
\end{figure} 
In order to calculate the nonfactorizable contribution, one must
first find a value for the
large energy light quark bozonisation coupling  $G_A$.
This was done  \cite{LLJOE} by requiring that our model should be 
consistent with
the equations (\ref{HECurrent}) and  (\ref{eq:charlesEq}). 
Applying the Feynman rules of LE$\chi$QM \cite{LLJOE} 
 we obtain the following bosonized current (before 
soft gluon emission forming a condensate is taken into account):
\begin{eqnarray}
J_0^\mu (H_{v_b} \rightarrow M_n) =   -N_c \int \dbar k
  \textrm{Tr} \left\{ 
\gamma^\mu L 
\,i S_v(k) 
\left[-iG_H H^{(+)}_{v_b} \right]
\,i  S_\chi(k) \,  
\left[i E \, G_A \, \gamma \cdot n \, Z \right] \,i S_n(k) \, 
  \right\} \, ,
\label{eq:M0start}
\end{eqnarray}
where  $\dbar k \, \equiv \, d^D k/(2 \pi)^D $ ($D$ being the dimension of
 space-time),
 and 
\begin{eqnarray}
S_v(k) \, = \, \frac{P_+(v)}{v \cdot k} 
\quad , \; S_\chi(k) \, = \, 
\frac{(\gamma \cdot k + m)}{k^2 - m^2}
 \quad  , \, S_n(k) \, = \,
\frac{\gamma \cdot n}{N \, n \cdot k} \; \, ,
\label{propagators}
\end{eqnarray}
are the propagators for heavy quarks  described by (\ref{constLagr}), for 
 light constituent quarks, and (\ref{eq:LEETdelta-Lag}) for 
light energetic quarks.
The presence of the left projection operator $L$ in $Z$ ensures that we only
 get contributions from the left-handed part of the interaction 
in (\ref{eq:LEETdHLciQM}), that is, 
$Z  \longrightarrow  -\xi^\dag M_L \, L$. 
 The  contribution in (\ref{eq:M0start})
 corresponding to the  
 $B \rightarrow \pi$ current is illustrated by the lower part of
the diagram in Fig.~\ref{fig:BDpiF}.

Loop diagrams within LE$\chi$QM  depend on momentum integrals of the form
\begin{eqnarray}
K_{r s t} = \int  \, \frac{\dbar k}{(v \cdot k)^r \, (k \cdot n)^s \,
  (k^2 - m^2)^t} \; \, ,
\label{eq:Krst}
\end{eqnarray}
\begin{eqnarray}
K^\mu_{r s t} = \int  \,
 \frac{\dbar k \; k^\mu}{(v \cdot k)^r \, (k \cdot n)^s \, (k^2-m^2)^t}
  = K_{rst}^{(v)} v^\mu + K_{r s t}^{(n)} n^\mu \; \, . 
\label{eq:Kmu}
\end{eqnarray}
These integrals have the  important property that
 $K_{r s t}^{(n)}$ dominates over 
 $K_{r s t}^{(v)}$
 and $K_{r s t}$ with one power of $1/\delta$. In the present model, we choose
$\nu = m$ which is of order $\Lambda_{QCD}$. Thus the  constituent light quark 
mass $m$  is the equaivalent of  $\Lambda_{QCD}$ within our model.
Some details of the calculation of the $B \rightarrow \pi$ is given in
Ref. \cite{LLJOE}.  
 
 To calculate emission of soft
 gluons we have used the framework of Novikov et al. \cite{Nov}. 
In that framework the ordinary vertex containing the gluon field
$A_\mu^a$ will be replaced by the soft-gluon version containing the
soft gluon  field tensor $G^a_{\mu \nu}$:
\begin{equation} 
 i g_s t^a \Gamma^\mu \, A_\mu^a \; \rightarrow  \; - \, \frac{1}{2}
 \, g_s \, t^a \, \Gamma^\mu \; G^a_{\mu \nu} 
\frac{\partial}{\partial k_\nu}\,  .... |_{k=0} \quad  ,
\label{Vertex}
\end{equation}
where $k$ is the momentum of the soft gluon. (Using this framework one has
 to be careful with the  momentum routing 
 because  the gauge where 
$x^\mu \, A^a_\mu =0$ has been used.)
 Here $\Gamma^\mu = \gamma^\mu \; , v^\mu $,
 or $n^\mu \, (\gamma \cdot \tilde{n} \, + \, \delta)/N$
for a light soft quark, heavy quark, or light energetic quark,
respectively. Our loop integrals are a priori depending on the 
gluon momenta $k_{1,2}$ which
are sitting in some propagators. These gluon momenta  disappear after
 having used the
procedure  in (\ref{Vertex}). (Note that the derivative has to be 
taken with respect to the whole loop integral).

Emision from the heavy quark or light
energetic quark are expected to be suppressed. This will be realized
in most cases because the gluon tensor is antisymmetric, and therefore such
contributions are often proportional to
\begin{equation} 
G^a_{\mu \nu} v^\mu \, v^\nu \; = \; 0 \quad , \quad {\mbox or} \quad 
G^a_{\mu \nu} n^\mu \, n^\nu \; = \; 0 \; .
\label{SupprGcond}
\end{equation}
However, there are also contributions proportional to  :
\begin{equation} 
G^a_{\mu \nu} v^\mu \, n^\nu \; \neq \; 0  \quad . 
\label{NonSupprGcond}
\end{equation}
analogous to what happens in some diagrams for the 
Isgur-Wise diagram where there are two different 
velocities $v_b$ and $v_c$ \cite{KresJ}.
Such  contributions appear within our calculation when two soft
 gluons are emitted from the heavy quark line.

 Using the prescription \cite{pider,BEF,ahjoe,Nov}
\begin{equation}
 g_s^2G^a_{\mu\nu}G^a_{\rho\lambda} \rightarrow 4\pi^2
\langle\frac{\alpha_s}{\pi}G^2\rangle                                    
\frac{1}{12}(g_{\mu\rho}g_{\nu\lambda}-g_{\mu\lambda}g_{\nu\rho}),
\label{GlueCorrel}
\end{equation}
for the gluon condensate one obtains the 
 leading bosonized current \cite{LLJOE}
\begin{equation}
J_{tot}^\mu(H\rightarrow M) =
 - i \frac{G_H \, G_A}{2} \,m^2 \, F \, 
   \textrm{Tr} \left\{ 
\gamma^\mu L 
H^{(+)}_v 
\left[\gamma \cdot n \right]
\xi^\dag M_L \right\} \; ,
\label{eq:Jmutot}
\end{equation}
where the quantity $F$ obtained from loop integration is a priori containing
a linearly divergent integral, which is related to the axial coupling $\ga$,
and can be traded for $\ga$. 
One  obtains \cite{LLJOE} for the quantity $F$: 
 \begin{equation}
F \; = \;  
\frac{3 \, f_\pi^2}{8 m^2 \, \rho} (1- g_A)  \, + \,  \frac{N_c }{16 \pi} \;
 - \,  \frac{(24 - 7 \pi)}{768 \, m^4} \, \gc \; \, .  
\label{FExpr}
\end{equation}
Note that $F$ is dimensionless.
The parameter $\rho$ is given in (\ref{GHcoupling}). 
Numerically, it was  found \cite{LLJOE} that $F \simeq 0.08$.

In order to obtain
the HL$\chi$PT Lagrangian terms $Tr(\bar{H}^aH^bv_\mu\mc{V}^\mu_{ba})$ and 
$Tr(\bar{H}^aH^b\gamma_\mu\gamma_5\mc{A}^\mu_{ba})$,  
having  coefficients $+1$ and $-g_\mc{A}$ respectively, one calculates 
quark loops with attached heavy meson fields
and vector and axial vector fields $\mc{V}^\mu$ or $\mc{A}^\mu$. 
Then  logarithmic and linearly divergent integrals
 obtained within the loop diagrams are identified with 
physical quantities or quantities of the model \cite{BEF,ahjoe,chiqm,pider}.

In order to fix $G_A$ in (\ref{eq:LEETdHLciQM}), we compare 
(\ref{HECurrent}) with (\ref{eq:Jmutot}).
In our  case where no  extra  soft  pions are going out,
  we put $\xi \rightarrow 1$, 
and for the momentum space $M_L \rightarrow k_M\sqrt{E}$, 
with the isospin factor
 $k_M = 1/\sqrt{2}$ for $\pi^0$ (while $k_M=1$ for charged pions).
 Moreover for the  $B$-meson
with spin-parity  $0^-$ we have
$ H^{(+)}_v 
\rightarrow P_+(v)(-i\gamma_5)\sqrt{M_H}$. Using this,
 the involved traces are easily calculated, and we obtain 
$J_{tot}^\mu(H\rightarrow M)$ for the case
 $\overline{B^0_d} \rightarrow \pi^+$:
\begin{equation}
J_{tot}^\mu(\overline{B^0_d} \rightarrow \pi^+) \, = \, 
 \frac{G_H G_A}{2} (\sqrt{M_H \, E}) \, m^2 \, F \, n^\mu.
\label{eq:MtotFTW}
\end{equation}
Using the equations (\ref{HECurrent}), (\ref{FExpr}), and 
(\ref{eq:MtotFTW}), one  obtains \cite{LLJOE}
 \begin{equation}
G_A \; = \; 
\frac{4 \zeta^{(v)}}{m^2 \, G_H \, F} \, \sqrt{\frac{E}{M_H}} \; ,
\label{GAExpr}
\end{equation}
where $\zeta^{(v)}$ is numerically known \cite{PBall}. 
 Within our model, the analogue of $\Lambda_{QCD}$ is the constituent
 light quark mass $m$.
To see the behaviour of $G_A$ in terms of the energy $E$, the quantity 
 $C$ in (\ref{eq:charlesEq}) is written as 
$C \, \equiv \, \hat{c} \, m^{\frac{3}{2}}$, which gives
\begin{equation}
G_A \; = \;
 \left(\frac{4 \hat{c} f_\pi}{m \, F \, \sqrt{2 \rho}} \right)
\;  \frac{1}{E^{\frac{3}{2}}} \; \, ,
\label{GAExpr2}
\end{equation}
which explicitly displays the behaviour $G_A \sim E^{-3/2}$. In terms of the
 number $N_c$ of colors, $f_\pi \sim \sqrt{N_c}$ and $F \sim N_c$ which gives
 the behaviour $G_A \sim 1/\sqrt{N_c}$, i.e. the same behaviour as the 
coupling  $G_H$ in (\ref{HLInteract}).

The bosonized current in (\ref{eq:Jmutot}) can now be written as 
\begin{equation}
J_{tot}^\mu(H\rightarrow M) =
 - 2 i \zeta^{(v)} \, \sqrt{\frac{E}{M_H}} 
 \; \,   \textrm{Tr} \left\{ 
\gamma^\mu L 
H^{(+)}_v 
\left[\gamma \cdot n \right]
\xi^\dag M_L \right\} \; .
\label{eq:JmutotBos}
\end{equation}

\section{Nonfactorizable Processe in LE$\chi$QM}

\begin{figure}[t]
\begin{center}
   \epsfig{file=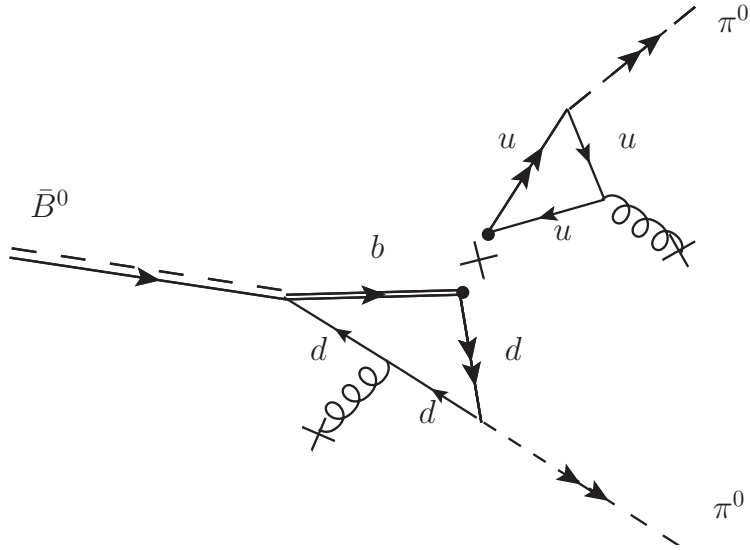,width=11cm}
\caption{Nonfactorizable contribution containing large energy
          light fermions and mesons. There is also corresponding diagram where 
the outgoing anti-quark $\overline{u}$ is hard.}
\label{fig:BDpiNF}
\end{center}
\end{figure}

In this section we calculate the nonfactorizable contribution to 
$\overline{B^0_d} \rightarrow \pi^0 \pi^0$ in Eq. (\ref{BDpiNonFact}).
 This will be
formulated as a quasi-factorized product of two coloured currents,
as illustrated in  Fig. \ref{fig:BDpiNF}.
Then the nonfactorized aspects enters through color correlation
between the two parts, using Eq. (\ref{GlueCorrel}).
Such a calculation within HL$\chi$QM and HL$\chi$PT is done
 previously \cite{EFHP}
for $\overline{B^0_{d,s}} \rightarrow  D^0 \overline{D^0}$. 
Here we will use the colored current for  $B \rightarrow \pi$, within
the LE$\chi$QM presented in the preceding section;
 see the diagram in Fig. \ref{fig:BDpiNF}.
 Using the $G_A$ value from the preceeding section, we may now 
calculate the nonfactorizable contribution to the process by adding 
one soft gluon to
each loop. Then we calculate the decay width
for  $\overline{B^0_d} \rightarrow \pi^0\pi^0$ from this
 nonfactorizable amplitude, 
and compare our results with experiment.

For a low energy quark interacting with one soft gluon, one might in
 simple cases use 
 the effective propagator \cite{RRY,BEF}
\begin{equation}
 S^G_1(k) \, = \, \frac{g_s}{4} t^a G^a_{\mu\nu}
\frac{(2 m \sigma^{\mu\nu}  \, + \left\{\sigma^{\mu\nu} , \gamma \cdot
  k \right\} )}{(k^2-m^2)^2} \; \, ,
\end{equation}
where $\left\{a,b\right\} \equiv a b + b a$ denotes the anticommutator.
This expression is constistent with the prescription in
(\ref{Vertex}),
and can be used for the diagram in Fig. \ref{fig:BDpiNF} .

Then one gets \cite{LLJOE} the following contribution to the
 bosonized colored $B
\rightarrow \pi$ current, shown in the lower part of
 the  diagram in Fig. \ref{fig:BDpiNF} :
\begin{eqnarray}
  J^\mu_{1G}(H\rightarrow M)^a \,  = \,  
    -\int \dbar k  \mathrm{Tr} \left\{ \gamma^\mu L t^a 
\, i S_v(k) \,                          
\left[-iG_H H^{(+)}_v \right]
\, i S_1^G(k) \, 
                        \left[i \, EG_A \, \gamma \cdot n Z \right]
\, i S_n(k) \; 
\right\} \, ,
 \label{eq:M1}
\end{eqnarray}
where $a$ is a color octet index.
Once more, we deal with the momentum integrals of the types
in (\ref{eq:Krst})  and (\ref{eq:Kmu}).
 Taking the color trace,  rewriting (\ref{eq:M1}), we obtain a
 contribution of the form
\begin{eqnarray}
  J^\mu_{1G}(H_b \rightarrow M)^a \, = \,
        g_s \, G^a_{\alpha \beta}  T^{\mu; \alpha \beta}(H_b \rightarrow M) \; ,
\label{eq:JtoM}
\end{eqnarray}
where the contribution from the 
(lower part of)
 the  diagram in Fig. \ref{fig:BDpiNF} :
alone is  to leading order in $\delta$
\begin{eqnarray}
  T^{\mu; \alpha \beta}(H_b \rightarrow M) \, = \,
          \frac{G_H \, G_A}{128 \pi}
 \epsilon^{\sigma \alpha \beta \lambda} \, n_\sigma
\mathrm{Tr} \left( \gamma^\mu L H^{(+)}_v
  \gamma_\lambda \,
    \xi^\dag M_L \right) \; ,
\label{eq:HtoM4}
\end{eqnarray}
where $E \cdot \delta = m$ has been explicitly used.

There are also other diagrams  not shown.  In one case the gluon is emitted from
the energetic quark. This diagram is zero due to (\ref{SupprGcond}).
Furthermore, there is a diagram not shown where the gluon is emitted from
the heavy quark which contains a non-zero part due to
(\ref{NonSupprGcond}). 
This gives an additional contribution to the colored $B \rightarrow \pi$
current
 which is nonzero.  However, this one will be
 projected out because it should be proportional to the Levi-Civita
 tensor to give a nonzero result for the
 $\overline{B_d^0} \rightarrow \pi^0  \pi^0$ amplitude as a whole, as will 
be seen from Eq. (\ref{eq:D0}) below.

The colored  current for an outgoing $\pi^0$ should  now be calculated
 in the LE$\chi$QM
(see upper part of the diagram in Fig. \ref{fig:BDpiNF}),
and we find
\begin{eqnarray}
  J^\mu_{1G}(M_{\tilde{n}})^a \,  = \,  
    -\int \dbar k  \mathrm{Tr} \left\{ \gamma^\mu L t^a 
 \, i S_1^G(k) \, 
 \left[i \, EG_A \, \gamma \cdot \tilde{n} \, Z \right]
\, i S_{\tilde{n}}(k) \; 
\right\} \, ,
 \label{eq:M2}
\end{eqnarray}
This colored $\pi^0$ current has the general form
\begin{equation}
  J^\mu_{1G}(M_{\tilde{n}})^a
\, = \,   g_s G^a_{\alpha \beta} \; 
 T^{\mu; \alpha \beta}(M_{\tilde{n}}) \; \, , 
\label{eq:D0cur}
\end{equation}
where the tensor $T$ is given by
\begin{equation}
    T^{\mu; \alpha \beta}(M_{\tilde{n}}) \, = \, 
   2 \, \left(-  \frac{G_A E}{4} \right) \, Y \, 
     \tilde{n}_\sigma\epsilon^{\sigma \alpha \beta \mu} \,   
 \mathrm{Tr}\left[\lambda^X \,  M_{\tilde{n}} \right] \; \; , 
\label{eq:D0}
\end{equation}
where the 
 $\lambda^X$ within the trace is the 
appropriate Gell-Mann  SU(3) flavor matrix. For an outgoing  hard $\pi^0$
this trace has the value $\sqrt{E/2}$ when going to the momentum space.
  The explicit factor 2 in front
 of this expression comes from the corresponding diagram, where
in the upper part of the diagram  
the antiquark could be hard and  the quark could be soft and
 emit a soft gluon. 
The factor $Y$ contains the result of loop momentum integration.
The relevant loop integral is now
\begin{eqnarray}
K^\mu_{0 1 2} = \int  \,
 \frac{\dbar k \; k^\mu} {(k \cdot n) \, (k^2-m^2)^2}
  = \frac{I_2}{\delta^2} \, n^\mu \; \, , 
\label{eq:K012mu}
\end{eqnarray}
which gives
\begin{equation}
    Y \,  = \,  - i I_2 \;  = \; 
\frac{ f_\pi^2}{4 m^2 N_c} \,\lambda \;  \equiv  \; Y_\lambda   
\, \equiv \, \frac{1}{4m^2 N_c} \left( f_\pi^2 \, - \,  
\frac{1}{24m^2}\langle\frac{\alpha_s}{\pi}G^2\rangle \right) \; .
\label{Q-factor}
\end{equation}
Here the parameter $\lambda$ 
 is of 
order $10^{-2}$ to $10^{-1}$ and very sensitive to small variations in
 the model dependent parameters $m$ and $\gc$.

It is easily seen that the experimental value of the 
$ \overline{B^0_d} \rightarrow \pi^0\pi^0$ amplitude can be  accomodated
for a constituent mass $m$ around 220 Mev and a value for $\gc^{1/4}$ 
around 315 MeV. These values are of the same order as used in previous articles
\cite{EFHP,ahjoeB,EHP,MacDJoe,LLJOE,EHFiz,ahjoe,EHbeta}.
But in contrast to these previous cases 
the present amplitude for $\overline {B^0_d} \rightarrow \pi^0\pi^0$ is very
 sensitive to variations of the model dependent parameters $m$ and $\gc$.
Or more specific, the colored  current   
$J^\mu_{1G}(M_{\tilde{n}})^a$ in 
(\ref{eq:M2}), (\ref{eq:D0cur}), and  (\ref{eq:D0}) is very 
sensitive to these parameters. In other
 words, $Y_\lambda $ has to be fine-tuned in order to produce the
 experimental result.

In a recent  paper \cite{KresJ} an extra mass parameter was introduced in the 
propagator of heavy quarks. One might do the same for propagator of 
the  light energetic 
quark, and use
\begin{equation}
S_n \; = \frac{\gamma \cdot n}{ N(n \cdot k + \Delta_n)} \; \, .
\end{equation}
 This would also bring this propagator more in harmony
 with the SCET propagator if $\Delta_n \sim p_\perp^2/E$ .  
This will to first order in $\Delta_n$ give an extra contribution in the
 loop integral obtained from the diagram in Fig.~2:
However, taking into account also the corresponding diagram where  the light 
anti-quark is the energetic one, this first order term in $\Delta_n$ cancels.
But there will be  terms of second order in $\Delta_n$, which are
 of order $\delta^2$.
Such contributions have to be considered together with higher order 
(in $\delta$) terms obtained from the interaction given by 
the operator $X$ in (\ref{eq:LEETdelta-Lag}).

One should note that the colored current given by  (\ref{eq:D0cur}) and 
(\ref{eq:D0})
 is determined 
by a triangle diagram. Thus one  might
 speculate if it can in some way be related  to the triangle anomaly. Namely,
the diagram in Fig.~\ref{fig:BDpiNF} would have,  
 for standard full propagators,  the mathematical properties of
 the diagram relevant for the triangle
 anomaly. Using dimensional regularization in this case, with dimension 
$D=4- 2 \epsilon$,  the loop integration 
gives an divergent result $\sim I_2 \sim 1/\epsilon$ while the 
corresponding Dirac trace is 
$\sim \epsilon$. Thereby one obtains  a
 finite expression for the triangle diagram in that case.
  However, in the present case
  we have replaced one of the standard (full)  quark propagators
by the SCET-like propagator $S_{\tilde{n}}$.  Then the  trace will not be 
$\sim \epsilon$ while  the corresponding loop integral is still
 divergent $\sim 1/\epsilon$. This  means  that the diagram is in total
 divergent.
Within our various chiral quark models including heavy quarks and
 light energetic quarks, the naive dimensional  
regularization (NDR) has been used, and divergent integrals have been 
identified with physical parameters
 \cite{EFHP,ahjoeB,EHP,MacDJoe,LLJOE,EHFiz,ahjoe,EHbeta,KresJ}. Using other
 schemes additional finite terms 
of type $\epsilon/\epsilon$ might appear \cite{BEF}, and some
parameters  might have  to be redefined.

We also note that the  description of the anomaly is rather tricky when
 going from the low
 energy process $\pi^0 \rightarrow 2 \gamma$ to higher energies where
  some cancellations occurr \cite{anomaly,EKP}.
In \cite{anomaly} the high energy processes $Z \rightarrow \pi^0 \gamma$ and 
  $\gamma^*  \rightarrow \pi^0 \gamma$ was studied. 
(Here the high energy  virtual photon 
$\gamma^* $ is coming from an energetic $e^+ e^-$ pair). 
In this case a part of the amplitude corresponding to low-energy decay
 $\pi^0 \rightarrow 2 \gamma$ is cancelled. But there is a remaining
 {\it anomaly tail} relevant for some high energy processes \cite{anomaly,EKP}.
 Trying to adapt such a description in our case, the tensor $T$
in (\ref{eq:D0}) for an outgoing $\pi^0$ and  soft gluon 
would be replaced by
\begin{equation}
    T^{\mu; \alpha \beta}(An) \, = \, 
    \frac{I_{An}}{4 \pi^2 f_\pi \sqrt{2}}  \, 
     p^\pi_\sigma  \, \epsilon^{\sigma \alpha \beta \mu} \,   
 \; , 
\label{eq:piAn}
\end{equation}
where we have taken into account that couplings and color traces are 
different from the calculations in \cite{anomaly,EKP}. 
The quantity $I_{An}$ is an integral given by 
\begin{equation}
    I_{An} \; = \; \int_0^1 \, \frac{x \, dx }{\eta \, x(1-x) -1}
 \; \; , 
\label{IAn}
\end{equation}
where $\eta \equiv p_\pi^2/m^2$.
Using as before $m$ as a constituent mass and $p^\pi = E \tilde{n}$ would give
$\eta = 1$ leading to $ I_{An} \simeq 0.6$. However, as the anomaly tail 
is of perturbative character \cite{anomaly,EKP} one  might think that it 
is more relevant 
to use masses closer to the current masses of order 5 to 10 MeV. In this
 case one has an assymtotic behaviour $ I_{An} \simeq ln(\eta)/\eta \;$, 
 and this would give values  for $ I_{An}$  of order $10^{-2}$.  

Now we use (\ref{GlueCorrel})
and also include the Fermi coupling
the Cabibbo-Kobayashi-Maskawa matrix elements, and the coefficient
 $2 c_A$ for the nonfactorizable
contributions to the amplitude, where $c_A$ is the Wilson coefficient
 for the $\mc{O}_A$ local operator. Using Eqs. (\ref{eq:JtoM}) and (\ref{eq:M2})
we find the effective Lagrangian at mesonic level  for the 
nonfactorizable contribution to $\overline{B^0_d} \rightarrow \pi^0\pi^0$
\begin{eqnarray}
  \mc{L}^{LE\chi QM}_{\textrm{Non.fact.}} \, = \, 
    \frac{4 \pi^2  c_A}{3}\left(4\frac{G_F}{\sqrt{2}}V_{ub}V^*_{ud}\right)
\gc \, S(H_b \rightarrow M_n \, M_{\tilde{n}}) \; ,
\label{eq:effLagrMes}
\end{eqnarray}
where $S(H_b \rightarrow M_n \,  M_{\tilde{n}})$ is the tensor product 
\begin{eqnarray}
S(H_b \rightarrow M_n \,  M_{\tilde{n}}) \; \equiv \; 
 T^{\mu; \alpha \beta}(H_b \rightarrow M_n) \;
  T_{\mu; \alpha \beta}( M_{\tilde{n}}) \; .
\label{eq:TensProd}
\end{eqnarray}

Using Eqs. (\ref{eq:HtoM4}) and (\ref{eq:D0}), and 
 $n \cdot \tilde{n} \simeq 2$ ,
we find the amplitude expressed entirely by known parameters, we find
 an explicite expression for $S(H_b \rightarrow M \, M_{\tilde{n}} )$
 in the case 
$\overline{B^0_d} \rightarrow \pi^0 \, \pi^0$: 
\begin{eqnarray}
S(\overline{B^0_d}  \rightarrow \pi^0 \pi^0) \; = \; 6 \, (\frac{1}{\sqrt{2}})^2
 \;  \frac{G_A^2 \,  G_H }{128 \pi}
\; Y \; 
\, E^2 \; 
   \sqrt{M_B} \; .                  
 \label{ExplicS}
\end{eqnarray}

We will now compare this nonfactorizable  amplitude for 
$\overline{B^0_d} \rightarrow \pi^0 \pi^0$ with the factorized amplitude which 
dominates $\overline{B^0_d} \rightarrow \pi^+ \pi^-$ :
\begin{eqnarray}
   \mathcal{M}_{\pi^+\pi^-} \, =
\, \left( 4 \, \frac{G_F}{\sqrt{2}}V_\textrm{ub}V^*_\textrm{ud} \right) \,
\cdot c_f \cdot 
\left(\frac{1}{2} \; J_\mu(\pi^-)
\right) \cdot 
\left(\frac{1}{2} J^\mu(\overline{B^0_d} \rightarrow \pi^+)
 \right)
\; ,
\label{BDpiFactAmp}
\end{eqnarray}
where
\begin{eqnarray}
 J_\mu(\pi^-) \, = \; f_\pi \, E \, \tilde{n}_\mu \; \; , \; \; 
 J^\mu(\overline{B^0_d} \rightarrow \pi^+) \; = \; 
  2 \, E \, n^\mu \, \zeta^{(v)} \; \, .
\label{Currents}
\end{eqnarray}
 The form factor  $\zeta^{(v)}$
is defined in 
(\ref{HECurrent}) and (\ref{eq:charlesEq}).  

Using the equations (\ref{GAExpr}) and  (\ref{eq:effLagrMes})- (\ref{Currents}),
 we find the following
 ratio between the non-factorized for $\overline{B^0_d} \rightarrow \pi^0\pi^0$
and the factorized amplitudes $\overline{B^0_d} \rightarrow \pi^+ \pi^-$ is
\begin{eqnarray}
r \; \equiv \;
 \frac{{\cal M}(\overline{B^0_d} \rightarrow \pi^0 \pi^0)_{\mbox{Non-Fact}}}
{{\cal M}(\overline{B^0_d} \rightarrow \pi^+ \pi^-)_{\mbox{Fact}}} \; = \; 
\frac{c_A}{c_f} \, \frac{\kappa}{N_c}
 \, \frac{ E \, \zeta^{(v)}}{ \sqrt{m \, M_B}} \; ,
\label{ratio}
\end{eqnarray}
where $\kappa$ is a  model-dependent hadronic factor 
\begin{eqnarray}
\kappa  \; = \; 
   \frac{\pi \, N_c \, \gc \, Y}{2  \, F^2 \, m^4 \sqrt{2 \rho}}
  \; .
\label{ExplicSEr}
\end{eqnarray}
It will  be interesting how the ratio $r$ scales with energy $E$. 
Using the scaling behaviour for $\zeta^{(v)}$
 with $C = \hat{c} m^{\frac{3}{2}}$ in (\ref{eq:charlesEq})
we find for the ratio $r$:
\begin{eqnarray}
r \; \simeq \;  
\frac{c_A}{ c_f} \, \frac{\kappa \,  \hat{c}}{N_c} \, 
\frac{m}{E} \; .
\end{eqnarray}
 Our calculations show that the ratio $r$ of the amplitudes are 
suppresed by $1/N_c$, as it should. 
The ratio is also scaling like $m/E$. Because $E \simeq m_b/2$ and $m$ is 
the equivalent of $\Lambda_{QCD}$ in our model, we have found that the 
nonfactorized amplitude is suppressed by $\Lambda_{QCD}/m_b$ as required 
by the analysis in ref. \cite{BBNS}.

\begin{figure}[t]
\begin{center}
   \epsfig{file=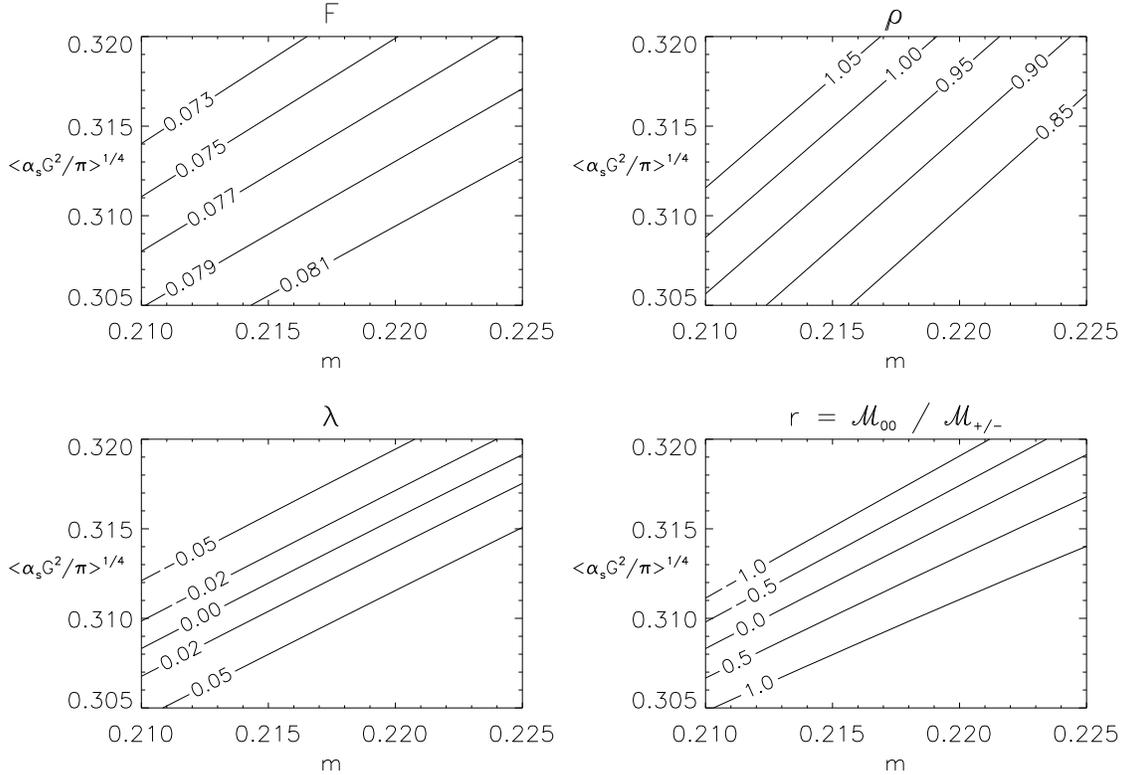,width=15cm}
\caption{Plots for  the quantities $F$, $\rho$, $\lambda$ and $r$
in terms of $m$ and $\gc^{1/4}$. We observe that for reasonable values
 of these parameters the ratio $r$ can take a wide range of values 
such that fine-tuning is required to reproduce the 
experimental value.}                   
\label{fig:Bpipi-plot}
\end{center}
\end{figure}

Concerning numerical predictions from our model, we have to 
stick to Eq. (\ref{ratio}). 
 The measured branching ratios for 
  $\overline{B_{d}^0} \rightarrow \, \pi^- \pi^{+}$
 and   $\overline{B^0_d} \rightarrow \pi^0 \pi^0$  are 
 $(5.13 \pm 0.24) \times 10^{-6}$ and 
$(1.62 \pm 0.31) \times 10 ^{-6}$, 
respectively \cite{PDG}. In order to predict the experimental value solely 
with the mechanism considered in this section, we should have 
$r \simeq 0.56 \pm 0.11$. Numerically, we use 
 $\zeta^{(v)} \, \simeq 1/3$ \cite{PBall}. 
In previous papers on the heavy-light chiral quark model
constituent masses 
$m \, \sim \, $ 220 MeV and $\gc^{1/4} \, \sim \, $  315 MeV
has been used. 
From the plot of $r$ in Fig.~\ref{fig:Bpipi-plot}, we observe that 
the experimental value of $r$ can easily be accomodated by values of 
such orders. The bad news is that
in our case the value of $Y_\lambda$ and thereby $\kappa$ and $r$ 
is very sensitive to the explicit choice of $m$ and $\gc^{1/4}$.
Thus fine-tuning has to be used. 

We also find that the perturbative anomaly tail will numerically reproduce the 
amplitude for $I_{An} \simeq 3.2 \times 10^{-2}$, corresponding to 
a quark mass $m_0 \simeq$ 11 MeV, i.e. of same order of magnitude as 
typical  current quark masses. Using a hybrid description with 
a quark model with constituent quark masses for the colored $\overline{B^0} 
\rightarrow \pi^0$ current in (\ref{eq:M1})-(\ref{eq:HtoM4}), and  the
 anomaly tail description \cite{anomaly,EKP} for the colored
 $\pi^0$ current in (\ref{eq:M2}), (\ref{eq:D0cur} and (\ref{eq:D0})
is not preferrable. Also, such a hybrid description also fails to 
show the behaviour $\Lambda_{QCD}/m_b$ required by QCD-factorization.
Still it might be  interesting that we can numerically 
match the colored current for outgoing $\pi^0$ with the anomaly
 tail description. 
.

\begin{figure}[t]
\begin{center}
   \epsfig{file=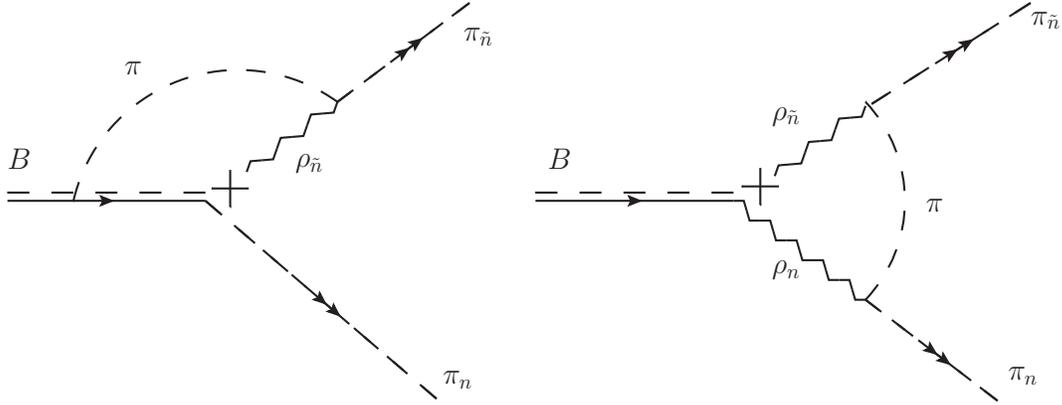,width=15cm}
\caption{Meson loops for $\overline{B^0_d} 
\rightarrow \pi \pi$. The zig-zag lines represent  energetic 
$\rho$-mesons. The dashed  lines with double arrow are energetic light mesons
and the dashed line with no arrow is a soft pion.}                   
\label{fig:meson-loops}
\end{center}
\end{figure}

Note that there are also mesonic loop contributions similar
 to those contributing to  
  processes  of the type $B \rightarrow \,
D \, \overline{D} \, $ and  $B \rightarrow \,
\gamma \, D \, $ \cite{EFHP,MacDJoe}. For those processes intermediate 
$D^*(1^-)$ mesons contributed. In the present case the analogous contributions 
would involve energetic vector mesons $\rho_n$, and we would need the
 amplitudes for $B \rightarrow \rho_n \rho_{\tilde{n}}$. Such loops are 
shown in Fig.~\ref{fig:meson-loops}. The diagram to the right would
 be calculable within an extended  theory involving energetic vector
 mesons. Unfortunately while the diagram to the left would be dubious 
because typical loop momenta would significantly exceed 1 GeV, and
 would require 
 insertion of  ad hoc form factors or should be handled
 within dispersion relation techniques. Both diagrams would of course require 
knowledge of the $\rho_n \, \pi_n \, \pi $ coupling in
  Fig.~\ref{fig:meson-loops}. In any case 
such calculations
 are beyond  the scope of this paper.

\section{Conclusion}

 We have pointed out  that the factorized 
amplitude for process
  $\overline{B_{d}^0} \rightarrow \,
\pi^0 \pi^{0} \, $ is proportional to a Wilson
 coefficient combination close to zero. Thus the nonfactorizable  
 contributions dominate the amplitude for this decay mode.
To handle the nonfactorizable contributions we have extended previous
 chiral quark models for the pure light quark case\cite{chiqm} used
 in \cite{pider,epb,BEF},  and the
 heavy light case\cite{ahjoe} used in
 \cite{ahjoeB,EHP,EFHP,EHbeta,MacDJoe,EHFiz},
 to include also energetic light quarks.

We have found that within our model we can account for  the 
amplitude needed to explain the experimental branching ratio 
for $\overline{B_{d}^0} \rightarrow \,
\pi^0 \pi^{0} \, $ \cite{PDG}. In addition the ratio $r$ between the 
non-factorizable and factorized amplitude scales as $\Lambda_{QCD}/m_b$
 in agreement with QCD factorization \cite{BBNS}.
However, the bad news is that the calculated amplitude is very sensitive to 
our model-dependent parameters, i.e. the constituent quark mass $m$, and the 
gluon condensate $\gc$. 
Anyway, final state interactions should be present\cite{KaidVysot}.

{\acknowledgments}

JOE is supported in part by the Norwegian
 research council
 and  in the beginning of this work, also by the European Commision RTN
network, Contract No. MRTN-CT-2006-035482 (FLAVIAnet).

\bibliographystyle{unsrt}

\end{document}